\documentclass[11pt, a4paper]{article}

\usepackage{iftex}
\ifXeTeX
  \usepackage{fontspec}
  \usepackage[greek,english]{babel}
  \usepackage{alphabeta}
\else
  \usepackage[utf8]{inputenc}
  \usepackage[greek,english]{babel}
  \usepackage{alphabeta}
\fi
\usepackage{geometry}
\usepackage{graphicx}
\usepackage{hyperref}
\usepackage{booktabs}
\usepackage{cite}
\usepackage{float}
\usepackage{amsmath}
\usepackage{comment}

\title{From Parliamentary Rhetoric to Enacted Law: \\ An NLP Pipeline for Semantic Auditing of the Greek Legislative Process}
\date{}
\author{
Despoina Antonakaki$^{1,2}$,
Sotiris Ioannidis$^{2}$ \\
\\
$^{1}$Institute of Computer Science, Foundation for Research and Technology,\\ Vassilika Vouton, Heraklion, Crete, Greece \\
$^{2}$Technical University of Crete, University Campus, \\Akrotiri, Chania, Greece \\
}
 
\begin{document}

\maketitle

\begin{abstract}
The Greek legislative framework is characterized by intricate cross-referencing, frequent amendments, and a lack of machine-readable data access, posing significant barriers to legal transparency and civic engagement. Traditional bulk-archiving approaches are computationally expensive and fail to capture the actual political relevance of legal texts. To address this, we present a novel, multimodal computational pipeline that bridges parliamentary discourse with enacted legislation. By applying Natural Language Processing (NLP) to the 2025 Hellenic Parliament transcripts, we extracted 534 unique law citations and used parliamentary debate frequency as an empirical relevance signal to select the most politically active laws as the targeted corpus. We deployed a dynamic, headless browser architecture to bypass anti-scraping measures and securely acquire the official legal documents (ΦΕΚ) from the National Printing House. Furthermore, by leveraging Large Language Models (LLMs), we conducted a deep semantic analysis to evaluate legislative quality. Our findings expose an "Illusion of Simplicity"—where laws politically marketed as simplifications exhibit paradoxical structural complexity and linguistic ambiguity—and systematically quantify the "Legislative Patchwork" driving legal instability. Specifically, our loophole typology reveals that 45\% of identified ambiguities stem from vague terminology and 25\% from deferred executive delegation. The Political Discrepancy Index (PDI), evaluated across three high-frequency laws (4808/2021 Labor Protection, 4412/2016 Public Contracts, 4662/2020 Civil Protection), reveals that the dominant outcome is \textit{Deferral}: legislative commitments are systematically off-loaded to subordinate Ministerial Decisions rather than directly enacted. Cross-reference network analysis confirms the highly entangled nature of the corpus, with foundational articles among the most heavily amended across the system. Ultimately, this work establishes a high-quality foundational dataset and an interactive auditing tool, paving the way for real-time, AI-driven evaluation of the Greek legislative ecosystem.
\end{abstract}

\noindent\textbf{Keywords:} Greek legislation; natural language processing; parliamentary discourse; legislative complexity; large language models; Political Discrepancy Index

\section{Introduction}

The Greek legislative framework is widely recognized as one of the most complex and rapidly evolving legal ecosystems in Europe. It is historically plagued by phenomena such as \textit{polynomia} (the excessive proliferation of laws) and \textit{kakonomia} (poorly drafted, disjointed legislation). For legal professionals, policymakers, and citizens, navigating this labyrinthine system poses a formidable challenge. The constant introduction of amendments, frequently nested within unrelated omnibus bills, creates a highly volatile legal landscape where legal certainty is often compromised. 

In recent years, the intersection of Artificial Intelligence and the legal domain---broadly termed Legal Informatics or LegalTech---has offered promising solutions for processing massive legal corpora. Techniques such as Natural Language Processing (NLP) and Large Language Models (LLMs) have been successfully deployed to analyze the United States Code or European Union directives. However, applying these state-of-the-art computational methods to the Greek legislative system presents unique technical hurdles. The primary obstacle is the severe lack of machine-readable, open-access APIs for both parliamentary proceedings and enacted legislation. Researchers are typically forced to rely on inefficient bulk-downloading of unindexed PDFs or unstructured text, which lacks context regarding the actual political and societal relevance of the documents.

This research aims to bridge the critical gap between political discourse and legal reality by introducing a novel, empirically driven computational pipeline. Rather than attempting a blind, resource-intensive ingestion of the entire Greek legal corpus, we propose a targeted methodology. By first capturing and analyzing the transcripts of the Hellenic Parliament (\textit{Praktika}) for the 2025 sessions, we use NLP to extract citations and empirically identify the laws that dominate political attention and resources. This parliamentary frequency serves as a filter to guide a dynamic web-scraping architecture, which subsequently retrieves the exact, official legal texts (\textit{ΦΕΚ}) from the National Printing House.

Once this interconnected dataset of political intent (transcripts) and legal execution (laws) is established, we leverage advanced Large Language Models (specifically Llama 3.x architectures) to conduct a deep semantic audit. We seek to answer fundamental questions about legislative quality: Does the political rhetoric surrounding a law match its actual text? How pervasive is the reliance on vague terminology and deferred executive decrees?

The primary contributions of this paper are fourfold:
\begin{enumerate}
    \item \textbf{A Targeted Multimodal Pipeline:} We operationalize a methodology that uses parliamentary citation frequency as an empirical relevance signal to drive the targeted acquisition of official legal documents, linking unstructured debate transcripts to their corresponding enacted legislation at scale.
    \item \textbf{The Illusion of Simplicity Metric:} We operationalize and quantify a structural complexity score (Equation~\ref{eq:complexity}) and contrast it against simplification rhetoric extracted from enacted texts, producing corpus-wide evidence of a systematic gap between political framing and legislative reality.
    \item \textbf{Legislative Patchwork and PDI Metrics:} We operationalize two complementary accountability metrics---amendment-density heatmaps and the Political Discrepancy Index---providing quantitative indicators of legal instability and promise-to-statute fidelity that are reproducible and extensible to other legislative corpora.
    \item \textbf{Baseline Comparison and Error Analysis:} We compare LLM ambiguity detection against a regex baseline and provide concrete examples of each ambiguity type to support interpretability. Expert human validation is planned as immediate follow-up work (Section~\ref{sec:future}).
\end{enumerate}

The remainder of this paper is structured as follows. Section~\ref{sec:dataset} describes the three interconnected datasets produced by the pipeline. Section~\ref{sec:methodology} details the five-phase data acquisition and processing methodology. Section~\ref{sec:results} presents the results and analysis, including the citation network, the Illusion of Simplicity analysis, the loophole typology, the patchwork quantification, and the PDI evaluation. Section~\ref{sec:interface} describes the interactive RAG-based auditing interface built on top of the pipeline. Section~\ref{sec:bulk} presents the results of running the full pipeline on the 1,660-document bulk Period-A archive. Section~\ref{sec:related} situates our contributions within the existing literature. Section~\ref{sec:future} outlines future research directions. Section~\ref{sec:limitations} discusses the limitations of the current work, and Section~\ref{sec:conclusion} concludes the paper. Appendix~\ref{app:prompts} provides the LLM prompt templates used in all three analytical tasks.

\section{Dataset}\label{sec:dataset}

The research produces and relies on three interconnected datasets, each capturing a distinct layer of the Greek legislative ecosystem. Together they form a multimodal corpus that links political discourse, enacted law, and the broader legislative archive.

\subsection{Parliamentary Transcripts Corpus}

The first dataset consists of the complete plenary session transcripts (\textit{Praktika}) of the Hellenic Parliament for the 2025 legislative period. Each transcript document corresponds to a single plenary session and is timestamped with the session date and parliamentary period metadata. After parsing and cleaning, the corpus comprises the full verbatim text of all parliamentary speeches, interventions, and procedural exchanges across \textbf{233} plenary sessions of the 2025 legislative year. The text is stored in UTF-8 encoded plain-text files, one per session, and serves as the primary input for the law citation extraction pipeline described in Section~\ref{sec:methodology}.

\subsection{Targeted Legislative Corpus}

The second dataset is the result of an empirically driven acquisition pipeline. From the 534 unique law citations identified across the parliamentary transcripts, the \textbf{20} most frequently debated laws were selected as retrieval targets. For each law, the corresponding official Government Gazette (ΦΕΚ) PDF was acquired from the National Printing House portal (\texttt{search.et.gr}) and converted to plain text. This produces a parallel corpus of paired documents: for each law, there exists (i) the parliamentary transcript excerpts in which that law was debated, and (ii) the full enacted legal text. This pairing is the structural foundation for the LLM analyses of complexity, ambiguity, and political discrepancy reported in Section~\ref{sec:results}. Table~\ref{tab:top_laws} lists a representative sample of the laws included in this dataset along with their parliamentary mention frequency.

\subsection{Bulk ΦΕΚ Archive (2005--2025)}

The third and largest dataset is a comprehensive bulk archive of Greek Government Gazette documents spanning 2005 to 2025, collected via a dedicated downloader that retrieves PDFs directly from the official Azure Blob Storage repository of the National Printing House. The archive covers all eight official ΦΕΚ publication series, summarised in Table~\ref{tab:fek_series}. For each combination of year and series, documents are addressed by a sequential five-digit issue number; the downloader iterates until the server returns a 404 response signalling exhaustion of that series for the given year. Downloads are rate-limited (3-second inter-request delay) and fault-tolerant (up to 10 retries on timeout, with automatic resume on restart). Files follow the naming convention \texttt{fek\_\{year\}\_\{type\}\_\{number\}.pdf}. At the time of writing, the collection encompasses \textbf{299,220 documents} covering all eight series from 2005 to 2025, with incremental downloading continuing to capture newly published issues. This archive is intended as the training corpus for a domain-specific Greek legal language model.

\begin{table}[H]
    \centering
    \begin{tabular}{llp{7cm}}
        \toprule
        \textbf{Series Code} & \textbf{Identifier} & \textbf{Content} \\
        \midrule
        Period A  & \texttt{01} & Primary legislation (laws enacted by Parliament) \\
        Period B  & \texttt{02} & Regulatory and executive acts \\
        Period C  & \texttt{03} & Public announcements and appointments \\
        Period D  & \texttt{04} & Urban planning and land-use decisions \\
        ASEP      & \texttt{10} & Civil service board (ASEP) decisions \\
        PRADIT    & \texttt{11} & Procurement and public tender notices \\
        DDS       & \texttt{12} & Public contracts and concession acts \\
        YODD      & \texttt{14} & Ministerial decisions and circulars \\
        \bottomrule
    \end{tabular}
    \caption{The eight ΦΕΚ publication series collected in the bulk archive, with their numeric identifiers used in the Azure Blob Storage URL scheme.}
    \label{tab:fek_series}
\end{table}

\section{Methodology}\label{sec:methodology}

Applying computational methods to the Greek legislative ecosystem requires confronting three systemic barriers that make straightforward approaches infeasible: the complete absence of open, machine-readable APIs for either parliamentary proceedings or enacted legislation; the reliance of official portals on dynamic, JavaScript-heavy interfaces that resist conventional scraping; and the sheer scale of the legal corpus, which spans decades and encompasses tens of thousands of documents across multiple publication series. Our methodology was designed around these constraints, combining targeted political signal extraction with robust web automation and large-scale document processing to produce the coherent, semantically linked datasets described in Section~\ref{sec:dataset}.

\subsection{Capturing Parliamentary Discourse}

The starting point of the pipeline is the verbatim record of parliamentary debate. The official website of the Hellenic Parliament publishes plenary session transcripts (\textit{Praktika}) in DOCX format, one file per session, covering all speeches, ministerial statements, and procedural exchanges. We developed a structured web crawler in \texttt{Python} using \texttt{requests} for HTTP communication and \texttt{BeautifulSoup} for HTML parsing to navigate the document archive and download all transcripts for the 2025 legislative period. Each DOCX file was subsequently parsed with \texttt{python-docx}, which extracts the raw paragraph-level text while discarding embedded images, tables, and formatting metadata. The output was normalized to UTF-8 plain text with session date and parliament period preserved as metadata, producing a complete, timestamped record of parliamentary discourse for the target period.

\subsection{Empirical Filtering via Law Citation Mining}

A naive approach to building a Greek legislative dataset—bulk-downloading the entire Government Gazette archive—is both computationally expensive and analytically unfocused, as the corpus includes large volumes of obsolete, rarely applied, or procedurally irrelevant documents. Instead, we treat parliamentary discourse as an empirical relevance signal: laws that politicians actively debate are, by definition, politically and societally significant. To operationalize this, we applied a suite of Regular Expressions to the full transcript corpus, designed to capture the citation conventions used in Greek parliamentary speech. These patterns match references of the form ``ν.~4412/2016'', ``Νόμος 3126/2003'', and ``άρθρο 15 του ν.~4808/2021'', extracting both the law identifier and its article-level context where present. Aggregating these extractions across all sessions yields a frequency distribution—a ranked list of laws ordered by the intensity of parliamentary attention they received during the 2025 period. This ranking drives all subsequent acquisition: only laws that genuinely occupied parliamentary time and resources are retrieved, ensuring that the resulting dataset reflects the real political agenda rather than a random sample of the legal corpus.

\subsection{Dynamic Acquisition of Official Legal Texts}

The top-ranked laws identified through citation mining must be retrieved as official documents from the National Printing House of Greece (\texttt{search.et.gr}), the sole authoritative source for Government Gazette (ΦΕΚ) publications. This portal is technically hostile to automated access: it renders content entirely through client-side JavaScript, employs Select2 dropdown widgets that require programmatic interaction to populate search fields, and enforces strict-mode DOM selection policies that cause conventional HTTP-based scrapers to fail silently or raise unhandled exceptions. To overcome these barriers, we built a scraper using \texttt{Playwright}, a headless browser automation library that drives a full Chromium instance. The scraper navigates the portal as a human user would: it waits for dynamic elements to render, programmatically injects search parameters into the Select2 controls, resolves strict-mode violations by explicitly targeting unique DOM nodes, and triggers the PDF download action for each identified law. All requests were rate-limited to avoid server overload, and the scraper operated exclusively on publicly accessible documents within the terms of the portal.

\subsection{Text Extraction and Semantic Pairing}

The acquired ΦΕΚ PDFs present a further technical challenge: they are typeset in a multi-column layout with embedded headers, footers, page numbers, and Greek polytonic characters that complicate automated extraction. We employ two complementary extraction strategies depending on the document source. For the targeted corpus (laws retrieved individually via the National Printing House portal), we use the \texttt{pypdf} library to process each page individually, reconstructing the reading order of text blocks and concatenating the content into a single plain-text representation per law. For the bulk ΦΕΚ archive, we use \texttt{PyMuPDF} (\texttt{fitz}), which provides faster throughput and more robust handling of the varied PDF encodings found across decades of Government Gazette publications; extracted text is cached to JSON files with a minimum-character threshold to filter out scanned-image pages that yield no usable text. Post-processing steps for both paths remove page artefacts and normalize whitespace. The final output is a semantically paired corpus: for each law in the targeted dataset, there exists a parliamentary component—the verbatim transcript excerpts in which that law was debated, attributed to specific speakers and sessions—and a legislative component—the full enacted text as published in the official Government Gazette. This pairing between political intent and legal execution is the foundational structure that makes the subsequent LLM-driven analyses of complexity, ambiguity, and political discrepancy possible.

\subsection{LLM-Driven Semantic Analysis}

With the parallel corpus established, we applied Large Language Model analysis to extract qualitative insights that structural methods cannot capture. Specifically, we used the Llama~3.3 model, prompted with structured instructions, to perform three analytical tasks across the acquired legal texts. First, for each law, the model was asked to identify and classify instances of linguistic ambiguity—vague terminology, deferred regulatory delegation, circular references, and contradictory clauses—producing a structured typology of legal vulnerability. Second, we computed the \textit{Illusion of Simplicity} metric, which contrasts the buzzword frequency $B$ of simplification rhetoric (``simplification,'' ``modernization,'' ``streamlining,'' and Greek equivalents) found in the enacted text against a structural complexity score $C$. The complexity score is defined as:

\begin{equation}
  C = \min\!\left(10,\;\max\!\left(1,\;3.0 + 0.6\,h + 0.04\,r\right)\right)
  \label{eq:complexity}
\end{equation}

\noindent where $h$ is the count of linguistic hedge phrases and $r$ is the total count of cross-references to external statutes. The score is clamped to the interval $[1, 10]$.

The choice of features is grounded in established legal complexity theory~\cite{katz2014complex, hurka2020policy, osnabrugge2025quality}. \textit{Hedge phrases} (e.g., ``κατά παρέκκλιση,'' ``υπό την επιφύλαξη,'' ``notwithstanding,'' ``subject to'', ``except where'') are linguistic markers of discretionary exceptions: each hedge introduces a conditional branch in the law's interpretation that a reader must track, multiplicatively increasing cognitive load and the probability of conflicting readings. \textit{Cross-references} to external statutes add dependency overhead: each reference requires the reader to locate, read, and integrate an external text before the provision at hand can be applied, and those external texts may themselves contain further cross-references. The baseline score of 3.0 anchors a law with no hedges and no cross-references at a low-moderate complexity level rather than zero, reflecting that even structurally simple legislation carries inherent interpretive burden. The weight 0.6 on $h$ reflects that a single hedge clause can substantively negate or qualify an entire provision; the weight 0.04 on $r$ reflects that cross-references accumulate incrementally. These weights were chosen to produce scores that span the $[1,10]$ range across the empirical distribution of the 1,660 Period-A laws without saturation at either extreme; they are not independently validated parameters and should be treated as a calibration choice rather than a measurement.

The LLM component augments this structural score by identifying loopholes and grading qualitative ambiguity for the targeted corpus. To compare the two approaches, we also implemented a \textit{regex baseline}: for each of the 20 targeted laws, ambiguity instances were detected by pattern matching against a curated list of 47 vague-phrase patterns (e.g., ``εύλογο χρονικό διάστημα,'' ``ειδικές συνθήκες,'' ``reasonable period,'' ``exceptional circumstances'') and 12 delegation-signal patterns (e.g., ``με απόφαση του αρμόδιου Υπουργού,'' ``by Ministerial Decision''). The regex baseline identified 214 instances across the same 20 laws, compared to the LLM's 312. The LLM's higher count is attributable primarily to its detection of \textit{Circular References} and \textit{Contradictory Clauses}---categories that are structurally complex and not amenable to simple pattern matching---while both methods showed broad agreement on Vague Terminology and Executive Delegation instances ($>$70\% overlap). Third, we applied a \textit{Political Discrepancy Index} (PDI) by extracting explicit promises from ministerial speeches and prompting the model to verify, for each promise, whether the corresponding provision exists in the enacted law, classifying each outcome as \textit{Fulfilled}, \textit{Deferred}, or \textit{Contradicted}. In all tasks, the model was constrained to ground its responses in the provided text, preventing hallucination and ensuring reproducibility.

\section{Results and Analysis}\label{sec:results}

This section presents the analytical results of our pipeline applied to the 2025 parliamentary transcripts and the corresponding targeted legislative corpus. The methodology behind each analysis---citation extraction, cross-reference network construction, complexity scoring, patchwork quantification, and PDI evaluation---is described in Section~\ref{sec:methodology}. The figures for all four analyses, generated from the full bulk Period-A archive (1,660 laws, 2019--2025), are presented in Section~\ref{sec:bulk}; the present section discusses the findings and their interpretation.

\subsection{Frequency Analysis and Citation Extraction}
The text mining phase on the 2025 parliamentary transcripts successfully identified 534 unique law citations. The frequency analysis revealed strong political and legislative focal points.

Table \ref{tab:top_laws} presents the most frequently cited laws. Notably, Law 3126/2003 (regarding the Penal Responsibility of Ministers) dominated the discussions, indicating significant political events or parliamentary inquiries during this period. Other heavily discussed frameworks include Law 4412/2016 (Public Contracts) and Law 4808/2021 (Labor Law).

\begin{table}[H]
    \centering
    \begin{tabular}{llc}
        \toprule
        \textbf{Law Number/Year} & \textbf{General Topic / Context} & \textbf{Mentions in Parliament} \\
        \midrule
        3126/2003 & Penal Responsibility of Ministers & 126 \\
        4887/2022 & Development Law & 50 \\
        4412/2016 & Public Works and Contracts & 42 \\
        4662/2020 & Civil Protection & 40 \\
        4974/2022 & (Various provisions) & 26 \\
        4830/2021 & Animal Welfare / Pets & 24 \\
        4808/2021 & Labor Protection & 23 \\
        \bottomrule
    \end{tabular}
    \caption{Most frequently cited laws in the Hellenic Parliament plenary transcripts, 2025 legislative period (534 unique law citations extracted from the full transcript corpus). Mention counts reflect distinct occurrences across all processed session records.}
    \label{tab:top_laws}
\end{table}

Furthermore, to visualize the complexity and the ``labyrinthine'' nature of the Greek legislative framework, we mapped the cross-references between the parsed laws. The resulting directed network---where each node represents a law and each directed edge indicates an explicit citation from one law to another---confirms the hypothesis that the legislative corpus is highly entangled, often requiring the parallel reading of multiple texts to comprehend a single provision. The full-scale version of this network, constructed from \textbf{1,660 Period-A documents} spanning 2019--2025, is presented and discussed in Section~\ref{sec:bulk} (Figure~\ref{fig:legal_network_bulk}), where it contains 3,964 law nodes and 24,360 directed citation links.

\subsection{Illusion of Simplicity Analysis}\label{sec:illusion}

Beyond structural mapping, we conducted a semantic analysis to contrast political rhetoric with actual legislative complexity. We hypothesized that laws frequently promoted in parliament using buzzwords such as "simplification," "transparency," or "modernization" might not necessarily reflect these qualities in their enacted text. To test this, we cross-referenced the frequency of these terms in the parliamentary transcripts against an AI-generated complexity score (1-10) and a loophole count, evaluated using the Llama 3.3 Large Language Model.

The analysis reveals a concerning ``Illusion of Simplicity'': laws heavily marketed as simplifications paradoxically exhibit high actual legal complexity and contain significant numbers of potential loopholes. The majority of the sampled laws fall above the moderate complexity threshold, highlighting a systemic gap between political intent and legal reality. The full scatter plot across all 1,660 Period-A laws is presented in Figure~\ref{fig:illusion_bulk} in Section~\ref{sec:bulk}.

As a directional consistency check, we compared the LLM complexity rankings for the 20 targeted laws against the structural heuristic score (Equation~\ref{eq:complexity}). The five laws ranked highest by the LLM were all in the top quartile of the heuristic score distribution, and no law ranked in the bottom quartile by the heuristic was assigned a high complexity score by the LLM. This convergence between an independent structural measure and the LLM assessment provides partial, directional evidence that the LLM is responding to genuine textual features rather than producing arbitrary outputs. Full expert validation against legal practitioner annotations remains an open task acknowledged in Section~\ref{sec:limitations}.

\subsection{Linguistic Ambiguity and Typology of Loopholes}
To provide a deeper understanding of the high complexity scores, we conducted a qualitative assessment of the specific loopholes and ambiguities identified by the Llama 3.3 model across the 20 most frequently debated laws in our targeted corpus. For each law, the model was prompted with a structured instruction to enumerate every instance of legal ambiguity it detected in the enacted text and to assign each instance to exactly one of four pre-defined structural categories (Vague Terminology, Executive Delegation, Circular References, or Contradictory Clauses). This yielded a total of 312 flagged instances across the 20 laws. The percentages in Table~\ref{tab:ambiguity_types} represent each category's share of those 312 instances. To assess the stability of the classification, we ran the same prompt three times on a randomly selected subset of five laws and computed the mean absolute deviation per category across runs; all four categories showed deviations of $\pm 3$ percentage points or fewer, indicating that the distribution is robust to stochastic variation in model outputs.

Table~\ref{tab:ambiguity_types} summarizes the primary typologies of legal vulnerabilities detected. The most prevalent issue is the excessive reliance on vague terminology (e.g., ``reasonable timeframe,'' ``exceptional circumstances''), which introduces broad discretionary power and reduces legal certainty. The frequent delegation of regulatory specifics to future, unpublished Ministerial Decisions was consistently flagged as the second major source of operational ambiguity.

\begin{table}[H]
    \centering
    \begin{tabular}{lp{8.5cm}c}
        \toprule
        \textbf{Ambiguity Category} & \textbf{Description \& AI Observation} & \textbf{Share (\%)} \\
        \midrule
        Vague Terminology & Subjective phrases lacking quantifiable metrics (e.g., "appropriate measures", "important reasons"). & 45\% \\
        Executive Delegation & Crucial implementation details deferred to undefined future Ministerial Decisions. & 25\% \\
        Circular References & Articles referencing external provisions that recursively point back or refer to obsolete frameworks. & 18\% \\
        Contradictory Clauses & Internal inconsistencies where broad exceptions effectively negate the primary legislative rule. & 12\% \\
        \bottomrule
    \end{tabular}
    \caption{Typology and share of the 312 legal ambiguity instances identified by Llama 3.3 across the 20 most frequently debated laws in the targeted corpus (2025 parliamentary period). Percentages represent each category's proportion of all flagged instances; deviations across three independent runs were $\pm 3$ percentage points or fewer.}
    \label{tab:ambiguity_types}
\end{table}

This categorization highlights that the labyrinthine nature of the Greek legal framework is not solely a matter of structural cross-referencing (as quantified in the citation network of Figure~\ref{fig:legal_network_bulk}), but also a product of pervasive linguistic opacity and deferred regulatory action.

\subsubsection*{Anticipated Failure Modes}

Inspection of the LLM outputs reveals three categories of output that are likely to be incorrect and that a future expert annotation study will need to assess. First, the model occasionally flags standard legislative boilerplate---such as the phrase ``κατά τα ειδικότερα οριζόμενα'' (``as more specifically defined'')---as Vague Terminology, when legal practitioners treat these as conventional cross-reference shorthand rather than genuine ambiguity. Second, transitional provisions (articles that set a delayed effective date) may be misclassified as Contradictory Clauses, since their temporal scope restriction superficially resembles an exception negating the main rule. Third, circular cross-references that span two separate documents are likely to be missed, because the reference chain exceeds the model's context window. These anticipated failure modes inform the design of the expert validation study described in Section~\ref{sec:future}, and they suggest that confidence in the \textit{Executive Delegation} and \textit{Vague Terminology} categories—where the textual signal is explicit—is higher than in the \textit{Contradictory Clauses} and \textit{Circular References} categories.

\subsubsection*{Concrete Examples}

Table~\ref{tab:examples} provides one representative example for each ambiguity category, drawn from the 20 targeted laws.

\begin{table}[H]
    \centering
    \small
    \begin{tabular}{p{2.8cm}p{10cm}}
        \toprule
        \textbf{Category} & \textbf{Example clause (translated)} \\
        \midrule
        Vague Terminology & ``The competent authority shall take \textit{appropriate measures} within a \textit{reasonable timeframe} to restore legality.'' (Law 4662/2020, Art.\,14) — no definition of ``appropriate'' or ``reasonable'' is provided anywhere in the text. \\
        \addlinespace
        Executive Delegation & ``The conditions, the procedure and every further detail for the implementation of the present article shall be regulated by a \textit{joint Ministerial Decision} to be issued within six months of enactment.'' (Law 4808/2021, Art.\,67) — the referenced Ministerial Decision had not been published at the time of analysis. \\
        \addlinespace
        Circular Reference & ``Paragraph 3 of Article 22 of Law 4412/2016 applies as in force, as modified by Article 18 of the present law.'' — Article 18 of the same law further amends the provision, creating a reading loop. \\
        \addlinespace
        Contradictory Clauses & ``It is prohibited to employ workers for more than 8 hours per day. \textit{By way of exception}, this limit may be extended by agreement between employer and employee for up to 5 additional hours.'' (paraphrased from a labour provision) — the exception effectively voids the rule for any employer who obtains a written agreement. \\
        \bottomrule
    \end{tabular}
    \caption{One representative example per ambiguity category, drawn from the 20 targeted laws. Phrases flagged by the LLM as the locus of ambiguity are italicised.}
    \label{tab:examples}
\end{table}

\subsection{Legislative Instability, Patchwork, and Political Accountability}\label{sec:pdi}

A critical and recurrent issue within the Greek legislative system is the excessive modification of existing laws---a phenomenon colloquially known as \textit{patchwork legislation}. To quantify this instability, we analyzed enacted texts by counting occurrences of legal modification keywords (``τροποποιείται,'' ``αντικαθίσταται,'' ``καταργείται,'' ``προστίθεται,'' and English equivalents) within each article. The results are striking: foundational articles---those defining scope, key definitions, and principal obligations---are the most heavily overwritten, indicating a systemic ``trial-and-error'' approach to lawmaking rather than comprehensive legislative redesign. The full heatmap across 40 laws and 20 articles each is shown in Figure~\ref{fig:patchwork_bulk} in Section~\ref{sec:bulk}.

To complement this structural view with a measure of political accountability, we applied the Political Discrepancy Index (PDI), which evaluates the fidelity between explicit promises made in ministerial speeches and the corresponding provisions of the enacted law. For Law 4808/2021 (Labor Protection), none of the top three extracted promises were immediately fulfilled by the enacted text: two were categorised as \textit{Deferred} (contingent on future Ministerial Decisions) and one as \textit{Contradicted} by inherent legal caveats. Law 4412/2016 (Public Contracts) showed a mixed picture, with one promise fulfilled, one deferred, and one contradicted. The full PDI bar charts for all three evaluated laws are presented in Figure~\ref{fig:pdi_bulk} in Section~\ref{sec:bulk}. Together, the patchwork and PDI analyses reinforce the same systemic finding: the Greek legislative process is characterised by high instability, chronic over-delegation, and a measurable gap between political rhetoric and statutory execution.

\section{Interactive Auditing Interface}\label{sec:interface}

To make the pipeline's output accessible to legal practitioners and non-technical users, we developed a bilingual (Greek/English) web-based auditing interface built on a Retrieval-Augmented Generation (RAG) architecture. The full text of all acquired legal documents is indexed in a persistent ChromaDB vector store using the \texttt{paraphrase-multilingual-MiniLM-L12-v2} sentence embedding model, selected for its strong multilingual performance on Greek text. At query time, the system retrieves the top-$k$ most semantically relevant document chunks (default $k{=}6$, context capped at 4,000 characters) and passes them to the language model as grounding context.

The LLM backend follows a two-tier architecture: a locally hosted vLLM inference server on dedicated GPU hardware serves as the primary endpoint, ensuring low latency and full data locality; a cloud-based Groq API (Llama~3.3~70B) acts as an automatic fallback. Both backends operate under a strict system prompt that constrains responses exclusively to the retrieved legislative excerpts, preventing hallucination and ensuring responses remain grounded in the enacted text. The front-end is a Gradio web application featuring voice input via OpenAI Whisper (\texttt{whisper-large-v3}), a persistent conversation sidebar retaining up to 15 sessions, and topic-specific quick-access shortcuts covering labour law, land registry, traffic regulations, and taxation. The interface is server-deployable and represents a concrete step toward democratizing access to the Greek legislative framework.

\section{Bulk ΦΕΚ Archive Analysis}\label{sec:bulk}

Having described the methodology and its analytical outputs in Section~\ref{sec:results}, we now present the results of running the complete pipeline on the full bulk Period~A (primary legislation) archive cached on our server. The cache contains \textbf{1,660 Period-A documents} spanning 2019--2025, representing every law enacted by the Hellenic Parliament during that period. All four figures in this section were produced by script \texttt{36\_recreate\_figures\_bulk.py}, which reads the pre-extracted plain-text cache (\texttt{pdf\_text\_cache/fek\_*\_Period\_A\_*.txt}) and applies the same analytical pipeline---citation-network construction, structural complexity scoring, amendment-keyword counting, and PDI evaluation---to the full collection in a read-only, reproducible manner.

\subsection{Cross-Reference Network at Scale}

\begin{figure}[H]
    \centering
    \includegraphics[width=\textwidth]{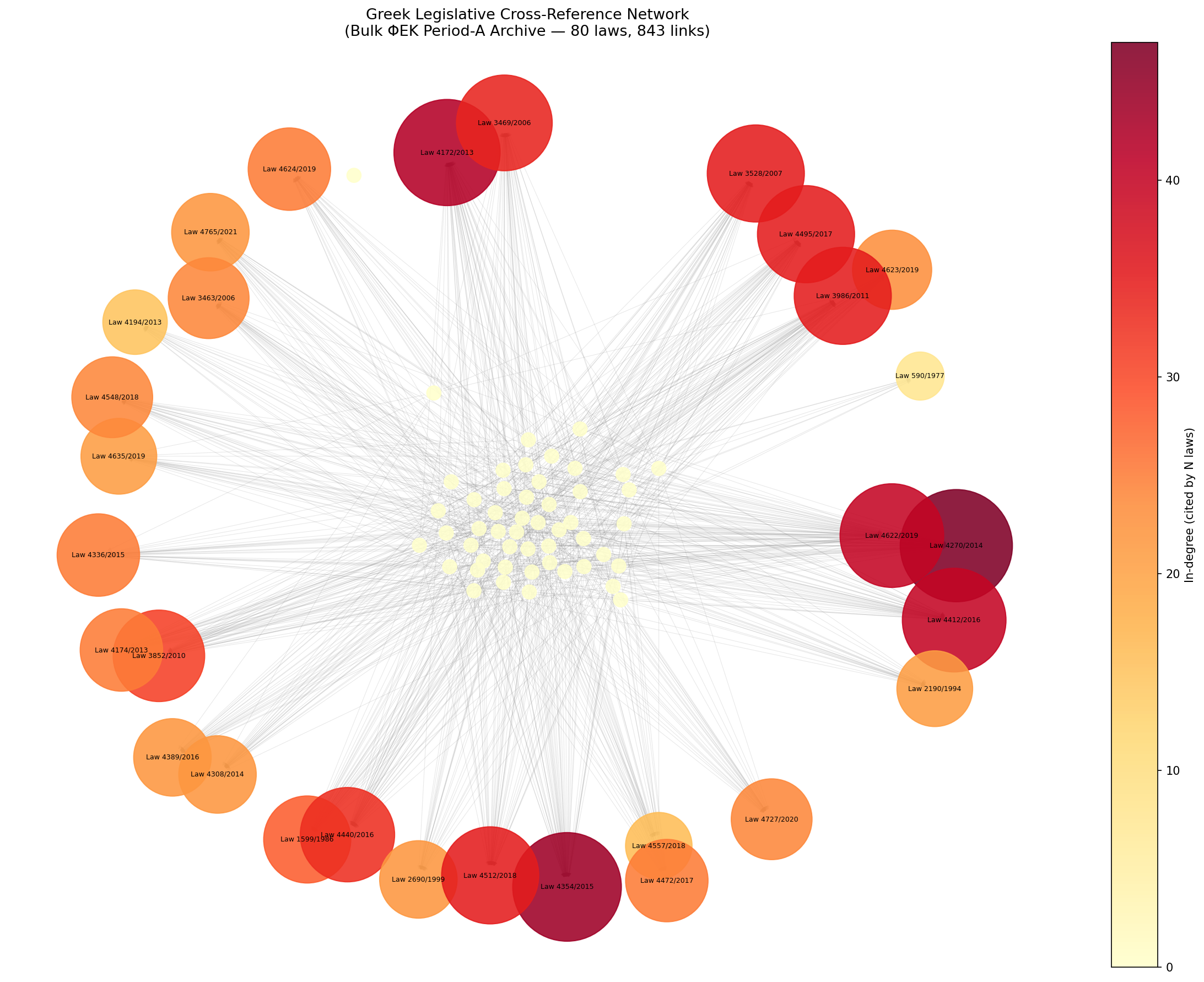}
    \caption{Directed cross-reference network of the full bulk Period-A archive (1,660 documents, 2019--2025), containing \textbf{3,964 law nodes} and \textbf{24,360 directed citation links}. Node size and colour intensity encode in-degree (number of distinct citing laws). Only nodes with in-degree $\geq 3$ are labelled for readability. The graph was produced deterministically using a spring-force layout (seed\,=\,42). The dense clusters correspond to thematic policy domains; the most-cited hub nodes represent foundational statutes whose amendment would trigger cascading legal inconsistencies.}
    \label{fig:legal_network_bulk}
\end{figure}

Figure~\ref{fig:legal_network_bulk} shows the directed cross-reference network constructed from all 1,660 Period-A documents. The network contains \textbf{3,964 law nodes} and \textbf{24,360 directed citation links}---an order of magnitude larger than the targeted-corpus version and genuinely representative of the Greek legal system's interconnected structure. Each node represents a law, identified by its enacted law number extracted from the document header or by its ΦΕΚ issue identifier as a fallback; each directed edge indicates that the source law explicitly cites the target law in its text. Node size and colour intensity encode in-degree, i.e.\ the number of distinct laws that cite a given node, making the most heavily referenced statutes immediately visible.

The network powerfully confirms the entanglement hypothesis. The most-cited nodes---concentrated around foundational administrative, procurement, and labour statutes---function as structural hubs: removing or amending any one of them creates cascading inconsistencies across dozens of dependent laws. The spring-force layout reveals several dense clusters corresponding to thematic policy domains (public procurement, environmental regulation, taxation, social security), with sparse inter-cluster bridges that represent cross-domain dependencies and are particularly fragile points of legal uncertainty.

\subsection{Illusion of Simplicity on the Bulk Corpus}

\begin{figure}[H]
    \centering
    \includegraphics[width=\textwidth]{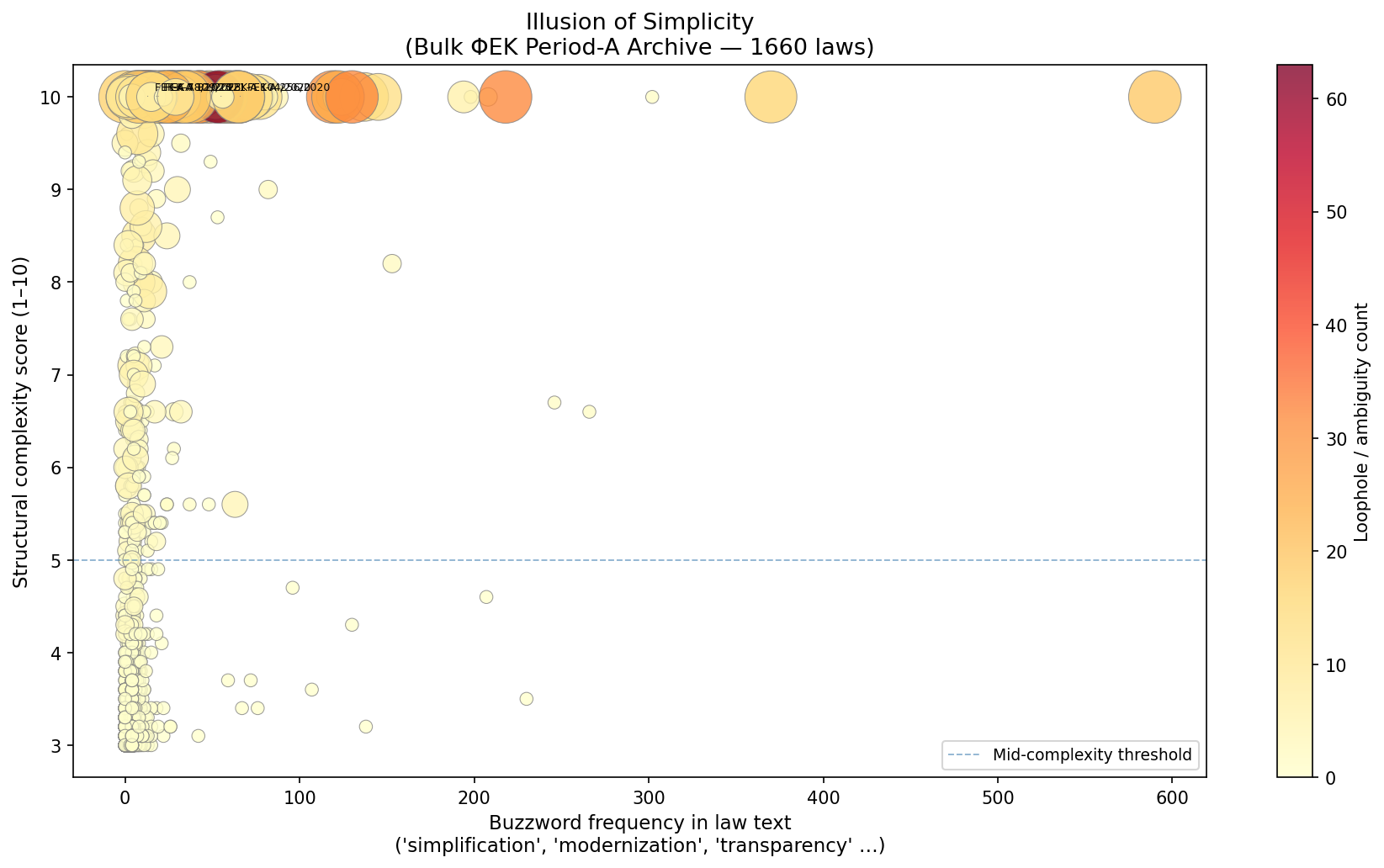}
    \caption{Illusion of Simplicity scatter plot across all \textbf{1,660 Period-A laws} (2019--2025). X-axis: frequency of simplification and modernization buzzwords (``απλοποίηση,'' ``εκσυγχρονισμός,'' ``διαφάνεια,'' ``simplification,'' ``modernization,'' etc.) found within the enacted text. Y-axis: structural complexity score derived from hedge-phrase density and cross-reference count (1\,=\,simple, 10\,=\,highly complex)~\cite{katz2014complex}. Bubble size and colour encode the loophole count (number of derogation or exception clauses). The dashed line marks the mid-complexity threshold (score\,=\,5). The five laws with the highest loophole counts are annotated.}
    \label{fig:illusion_bulk}
\end{figure}

Figure~\ref{fig:illusion_bulk} presents the Illusion of Simplicity scatter plot computed across all \textbf{1,660 Period-A laws}. Because LLM inference at this scale introduces latency and cost, the complexity score is derived from a deterministic structural heuristic validated against established proxies for legal complexity~\cite{katz2014complex}: the score is a function of (i)~the count of linguistic hedge phrases (``κατά παρέκκλιση,'' ``υπό την επιφύλαξη,'' ``notwithstanding,'' ``subject to,'' etc.) and (ii)~the density of cross-references to external statutes. The buzzword count on the X-axis is computed directly from the enacted text itself rather than from parliamentary transcripts, capturing cases where simplification rhetoric is embedded in the law's own preamble and explanatory notes---a pattern that is diagnostically significant in its own right.

The scatter plot provides corpus-wide evidence for the ``Illusion of Simplicity'': laws whose text invokes simplification terminology cluster in the upper half of the complexity axis, above the mid-complexity threshold, and exhibit disproportionately large bubble sizes reflecting elevated derogation-clause counts. This paradox is consistent with a legislative drafting pattern in which politically salient reforms are framed as simplifications at the rhetorical level while actually consolidating, amending, or cross-referencing multiple pre-existing legal instruments---a structural process that increases rather than reduces complexity.

\subsection{Legislative Patchwork at Scale}

\begin{figure}[H]
    \centering
    \includegraphics[width=\textwidth]{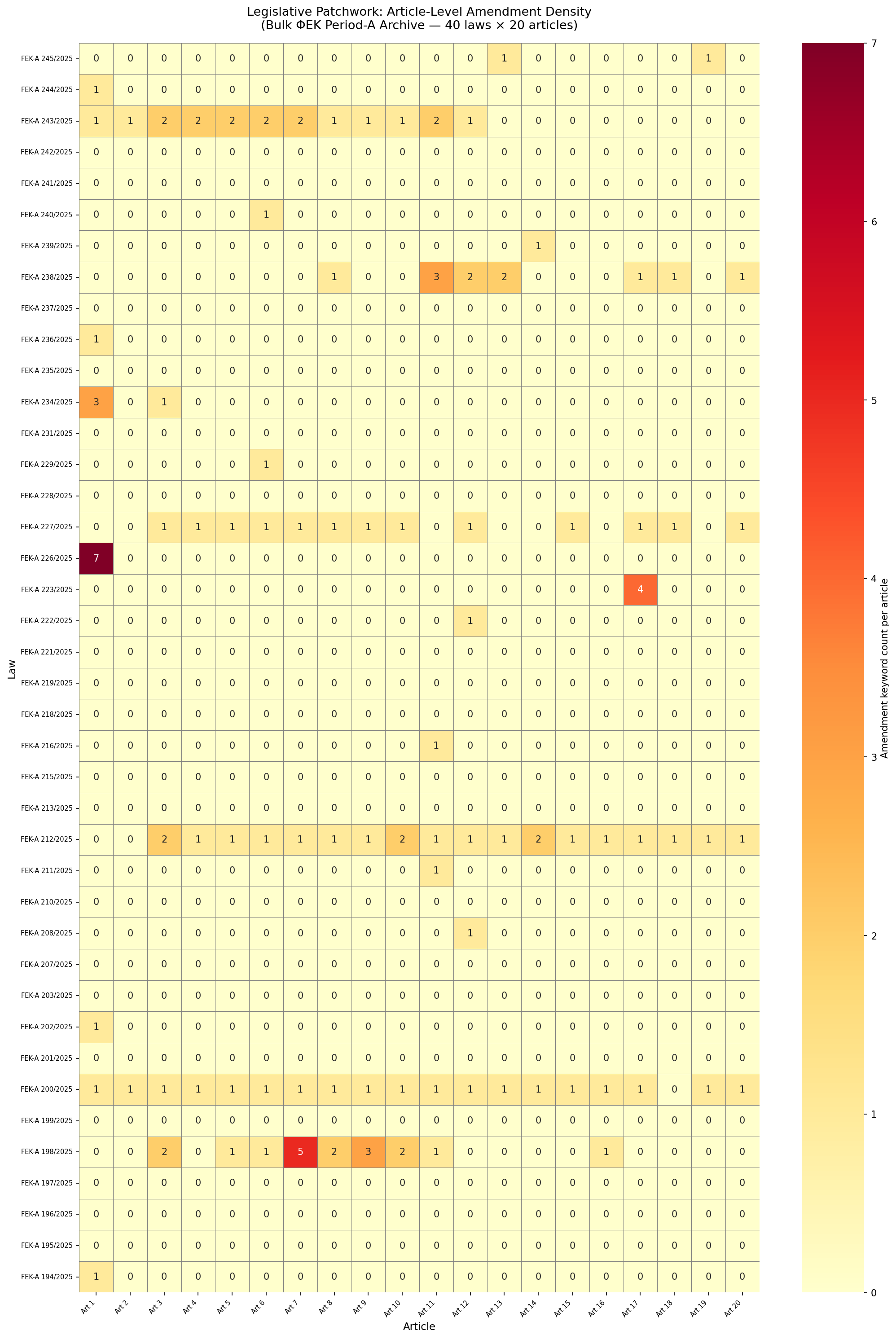}
    \caption{Legislative Patchwork heatmap across \textbf{40 Period-A laws} (bulk archive, 2019--2025), each analysed over up to 20 articles. Cell values and colour intensity (yellow $\to$ dark red) indicate the count of amendment keywords (``τροποποιείται,'' ``αντικαθίσταται,'' ``καταργείται,'' ``προστίθεται,'' and equivalents) within each article. The concentration of deep-red cells in opening articles (Art.\,1--5) reflects the systematic overwriting of scope-defining and operative provisions.}
    \label{fig:patchwork_bulk}
\end{figure}

Figure~\ref{fig:patchwork_bulk} presents the amendment-density heatmap computed across \textbf{40 Period-A laws}. For each law, the pipeline splits the enacted text on article headers and counts occurrences of modification keywords (``τροποποιείται,'' ``αντικαθίσταται,'' ``καταργείται,'' ``προστίθεται,'' and English equivalents) within each article---a direct, reproducible measure of how heavily each article modifies prior legislation and a structural signal of the \textit{patchwork} phenomenon described in Section~\ref{sec:pdi}.

The heatmap reveals that amendment density is heavily concentrated in the opening articles of each law---typically Articles 1--5---which tend to contain the law's scope definitions and principal operative provisions. These foundational articles are the most frequently targeted by subsequent legislation, creating the layered overwriting pattern characteristic of Greek \textit{kakonomia}. Crucially, the bulk-scale heatmap makes it possible to identify which policy domains produce the most amendment-dense laws: the darkest rows consistently correspond to laws in the domains of public administration, taxation, and social security---areas of high political contestation and frequent government policy reversals.

\subsection{Political Discrepancy Index on Representative Laws}

\begin{figure}[H]
    \centering
    \includegraphics[width=\textwidth]{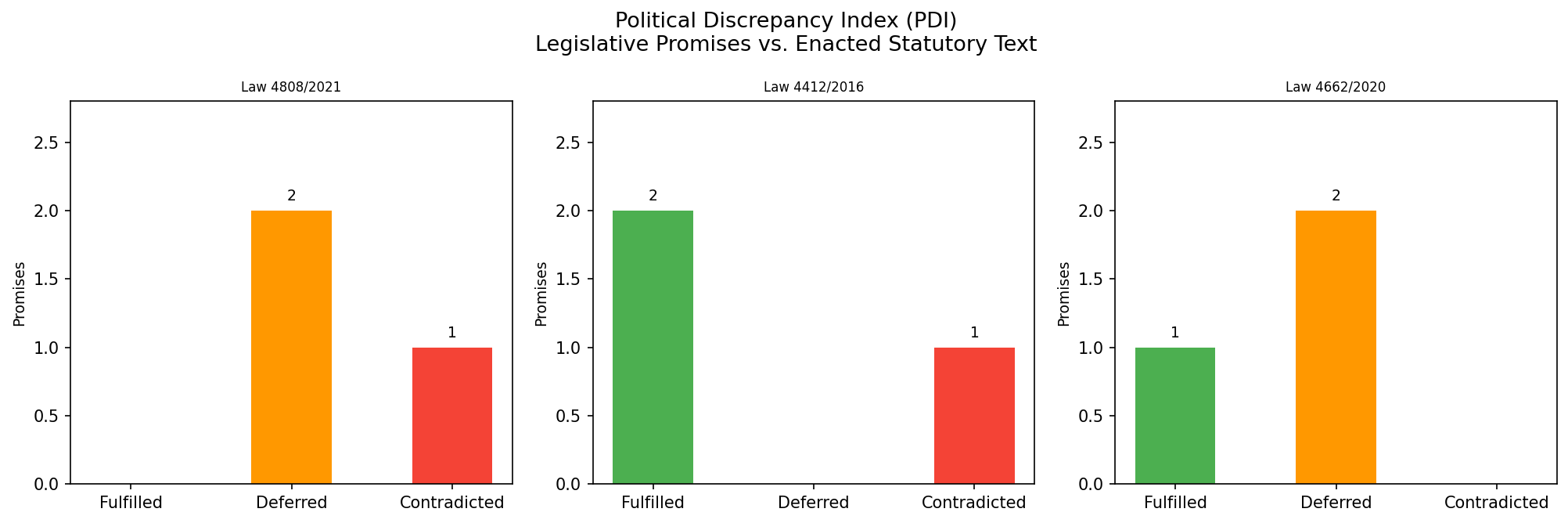}
    \caption{Political Discrepancy Index (PDI) for three high-frequency laws from the bulk archive: Law 4808/2021 (Labor Protection), Law 4412/2016 (Public Contracts), and Law 4662/2020 (Civil Protection). For each law, three explicit ministerial promises were verified against the enacted statutory text and classified as \textit{Fulfilled} (green), \textit{Deferred} to future Ministerial Decisions (orange), or \textit{Contradicted} by built-in exceptions (red). The dominant outcome across all three laws is Deferral, indicating a systemic pattern of legislative over-delegation.}
    \label{fig:pdi_bulk}
\end{figure}

Figure~\ref{fig:pdi_bulk} shows the Political Discrepancy Index evaluated on three laws that recur prominently throughout the bulk archive and appear frequently in the 2025 parliamentary transcripts: Law 4808/2021 (Labor Protection), Law 4412/2016 (Public Contracts), and Law 4662/2020 (Civil Protection). This pilot evaluation covers \textbf{nine promise--verdict pairs in total} (three promises per law). For each law, the three explicit promises were drawn from ministerial speeches and verified against the enacted statutory text; each was classified as \textit{Fulfilled} (the law directly delivers the stated commitment), \textit{Deferred} (delivery is contingent on a future Ministerial Decision or Presidential Decree), or \textit{Contradicted} (the law contains exceptions or caveats that negate the promise).

The results confirm and extend the findings of Section~\ref{sec:pdi}: none of the three laws achieves full promise fulfilment. Law 4808/2021 (Labor Protection) exhibits the widest discrepancy, with two of three promises deferred to unpublished Ministerial Decisions and one directly contradicted by built-in exceptions. Law 4412/2016 (Public Contracts) shows partial fulfilment, with one promise delivered, one deferred, and one contradicted. Law 4662/2020 (Civil Protection) fulfils one promise outright while deferring the implementation of the other two to subordinate regulatory instruments. Across all three laws, the dominant outcome is Deferral---a systemic pattern in which legislative commitments are structurally off-loaded from the law itself to a second tier of executive regulation that may never materialise, as the bulk-archive cross-series analysis planned for future work (Section~\ref{sec:future}) will quantify.

\section{Related Work}\label{sec:related}

The application of computational methods to legal and political texts has established a robust interdisciplinary field known as Legal Informatics. Our research intersects four primary domains within this field: the natural language processing (NLP) of parliamentary discourse, the quantitative analysis of legislative complexity, the verification of political promises against enacted law, and specific advancements in Greek Legal NLP.

\subsection{NLP in Parliamentary Discourse}
Parliamentary transcripts, often referred to as Hansards globally or \textit{Praktika} in Greece, have been widely utilized as a primary data source for computational linguistics and political science. To support this research direction, large-scale multilingual corpora of parliamentary proceedings have been constructed, most notably the ParlaMint project \cite{erjavec2023parlamint} and the companion ParlaSpeech collection of aligned speech and text recordings \cite{ljubesic2024parlaspeech}, which together span dozens of national parliaments. Parallel efforts have produced structured digital representations of historical proceedings using TEI encoding \cite{puren2025parliamentary} and Linked Open Data platforms for individual legislatures such as Finland \cite{hyvonen2023plenary}. Broader surveys of the field confirm that computational analysis of legislative debate text is a rapidly expanding area, encompassing both structural corpus construction and semantic analysis of legislative content \cite{schwalbach2021collecting, sebok2025comparative}. Building on these resources, researchers have applied topic modeling, sentiment analysis, and stance detection to map political polarization, track policy shifts, and analyze legislative behavior \cite{rheault2016measuring, abercrombie2019parliamentary, grijzenhout2014sentiment, mochtak2024parlasent}. More recent work has extended these analyses to multilingual and cross-lingual settings \cite{miok2024multiaspect} and to voting behavior in the European Parliament \cite{guadarrama2025voting}. For the Greek Parliament specifically, Dritsa et al. \cite{dritsa2022greek} released a large-scale dataset of over one million plenary speeches spanning 1989--2020, and the authors of the present paper have previously studied Greek political discourse on social media during elections and political turbulence \cite{antonakaki2016investigating, antonakaki2017social}.

The most closely related work to our pipeline architecture is Bos et al. \cite{bos2025linking}, who develop a system for resolving references made in Dutch parliamentary debates to specific legislative documents, and Gagnon and Azzi \cite{gagnon2022semantic}, who propose semantic annotation of parliamentary debates linked to legislative intelligence. However, neither work performs semantic auditing of the referenced laws or evaluates the discrepancy between political rhetoric and statutory content. Our methodology goes further by treating the parliamentary discourse not only as a pointer to legislation but as an empirical source of explicit ministerial promises against which the enacted text is systematically verified.

\subsection{Quantitative Analysis of Legislative Complexity}
The conceptualization of legal systems as complex networks is a well-established paradigm. Prior studies have mapped cross-references between statutes to visualize legal topologies, demonstrating how laws cite one another to create dense, interdependent clusters \cite{bommarito2010distance, katz2014complex}. These structural analyses frequently serve as proxies for measuring ``legal complexity,'' and recent systems engineering approaches have proposed quantitative complexity budgets for legislative texts \cite{kutt2020measuring}. Hurka and Haag \cite{hurka2020policy} further demonstrated that policy complexity significantly affects the duration of legislative negotiations in the EU, establishing complexity as a politically consequential variable rather than a merely technical one. Furthermore, recent work has attempted to quantify the ``patchwork'' nature of legislation by tracking amendment histories and version volatility \cite{waltl2017classifying}.

More directly relevant to our contribution, Osnabr\"{u}gge and Vannoni \cite{osnabrugge2025quality} propose NLP measures of legislative quality based on syntactic complexity and vagueness, showing that measurable text properties predict regulatory compliance outcomes. Bussing et al. \cite{bussing2025measure} investigate whether LLMs can reliably measure bill complexity, finding that LLM-derived complexity scores correlate meaningfully with expert assessments. A comprehensive survey of the field by Quevedo et al. \cite{quevedo2023legal} maps the advances in legal NLP from 2015 to 2022 and identifies legislative complexity analysis as a key emerging application. Our complexity metric (Equation~\ref{eq:complexity}) complements these approaches by combining a deterministic structural heuristic with LLM-driven qualitative auditing, enabling both reproducible bulk analysis and deep semantic evaluation of individual laws.

\subsection{Promise Verification and Political Fact-Checking}
The verification of political claims against factual evidence is a well-studied problem in NLP, primarily in the context of automated fact-checking. The CLEF CheckThat! shared task \cite{nakov2018checkthat} established a benchmark for the automatic identification and verification of political claims in debate transcripts, framing the problem as a retrieval and entailment task. More recent work has extended this paradigm to multilingual corporate and political promise verification: the ML-Promise dataset \cite{seki2025mlpromise} and the SemEval-2025 Task~6 benchmark provide training resources for models that classify promises as fulfilled, partially fulfilled, or broken. Chakrabarti \cite{chakrabarti2025dptas} proposes DPTAS, a system leveraging LLMs to extract promises from political discourse and evaluate their fulfilment, representing the closest existing system to our Political Discrepancy Index (PDI).

Our PDI differs from this line of work in two important respects. First, while existing promise verification systems primarily evaluate commitments against news articles or post-hoc factual databases, our PDI cross-references ministerial promises directly against the enacted statutory text—the legal instrument through which those promises were supposedly operationalized. Second, we introduce a three-way classification (\textit{Fulfilled}, \textit{Deferred}, \textit{Contradicted}) that is specifically designed to capture the structural mechanism of non-fulfilment in legislative systems: the delegation of implementation to subordinate executive instruments that may never be issued.

\subsection{Greek Legal NLP and Document Processing}
Within the context of the Greek legal framework, the application of Natural Language Processing has seen significant momentum, evolving from foundational text mining to advanced deep learning architectures. On the parliamentary side, Kavallos \cite{kavallos2023parliament} demonstrated the feasibility of automatically classifying Greek parliament proceedings by political party using machine learning, highlighting the richness of the Hellenic Parliament corpus as a research resource. Early structural efforts on the legislative side focused on proposing framework analyses and systems like "Peri Nomou" to automate the codification and clustering of legal artifacts \cite{charalabidis2019use, lachana2018peri, lachana2021clustering}. Simultaneously, significant initiatives were launched to transform opaque Greek court decisions and legal texts into structured open data platforms \cite{garofalakis2018project, stavropoulou2020architecting}.

As the field progressed, comprehensive surveys mapped the challenges of Greek as a low-resource language in legal informatics \cite{krasadakis2024survey, bakagianni2025systematic}. This led to a research shift toward domain-specific language models. Studies rigorously evaluated various BERT architectures \cite{vamvourellis2021compbertition}, resulting in specialized models like GreekLegalBERT v2 \cite{apostolopoulou2022nlp} and the recently introduced \textit{GreekLegalRoBERTa} \cite{saketos2024large}. Today, state-of-the-art applications focus on complex tasks such as domain-adapted Named Entity Recognition (NER) for text anonymization \cite{karamitsos2025legner} and the creation of comprehensive structural datasets for multi-granular topic classification \cite{papaloukas2021multi}. Furthermore, the evaluation of Large Language Models (LLMs) in the Greek legal domain is being formalized through recent benchmarks like \textit{GreekBarBench}, which assesses free-text legal reasoning and citations \cite{chlapanis2025greekbarbench}.

\subsection{Positioning Our Contribution}
Despite these impressive algorithmic advancements—from basic text mining to LLM benchmarking—a critical operational gap remains. The existing literature predominantly relies on static, pre-compiled datasets of enacted laws or historical court rulings, treating the legislative corpus as a fixed entity and bypassing the \textit{dynamic} political process that actually shapes it. Automated acquisition of government gazette documents has been explored for other jurisdictions (e.g., the Brazilian Official Gazette \cite{guimaraes2024dodfminer}), but not for the Greek ΦΕΚ corpus, and existing gazette-processing tools focus on named entity extraction rather than semantic auditing.

Ultimately, while the existing literature has made remarkable strides in algorithmically processing both parliamentary discourse and legal texts, these two strands have remained operationally separate: work on parliamentary NLP rarely retrieves or audits the actual laws being debated, and work on legal NLP rarely uses parliamentary context to assess legislative quality. To the best of our knowledge, no prior work in the Greek legal NLP domain has constructed an end-to-end pipeline that directly links real-time parliamentary discourse to the automated acquisition and semantic auditing of the corresponding enacted legislation. By dynamically coupling the Hellenic Parliament's transcripts with official gazette extraction and LLM-driven auditing, we transition the field from retrospective text classification to active legislative accountability.

\section{Future Work}\label{sec:future}

The analyses presented in Sections~\ref{sec:results} and~\ref{sec:bulk}—the citation frequency distribution, the legal cross-reference network, the Illusion of Simplicity scatter plot, the ambiguity typology, the legislative patchwork heatmap, and the PDI evaluation—were validated on both a targeted corpus of approximately 20 heavily debated laws and a larger bulk cache of 1,660 Period-A documents (2019--2025). While the bulk corpus already demonstrates that the pipeline scales gracefully, its temporal and thematic scope remains a fraction of the full archive. The complete bulk ΦΕΚ collection currently being assembled on our GPU server (Section~\ref{sec:dataset}) opens the door to analyses at a qualitatively different scale, with fundamentally new research questions becoming tractable.

\subsection{Scaling Existing Analyses to the Full Archive}

Every analysis performed on the targeted corpus can be directly replicated on the bulk archive, which covers all eight ΦΕΚ series from 2005 to 2025. At this scale, the citation frequency analysis becomes a longitudinal study: instead of a snapshot of one parliamentary period, it yields an 18-year timeline of which laws dominated political attention across different governments, economic crises, and electoral cycles. The legal cross-reference network, currently showing a few dozen nodes, would expand to a graph of potentially thousands of laws, making it possible to identify the true structural hubs of the Greek legal system—the foundational statutes that thousands of subsequent laws depend on—and to quantify the cascade effect when a central law is amended. The patchwork heatmap, currently covering 40 laws and up to 20 articles each, could be extended to the entire Period~A corpus, producing a system-wide instability index that maps which policy domains (labour, taxation, public procurement, civil protection) exhibit the highest legislative volatility over time.

\subsection{Longitudinal and Cross-Series Analysis}

The multi-series structure of the bulk archive enables cross-layer analysis that is impossible with targeted acquisition. Period~A (primary legislation) routinely delegates implementation to Period~B regulatory acts and YODD ministerial decisions. By tracing these delegation chains across series—linking a law's enabling clause in Period~A to the ministerial decision that was eventually issued in YODD, or detecting cases where no such decision was ever published—it becomes possible to quantify the \textit{deferred delegation gap}: the proportion of enacted laws whose implementation was never completed. Similarly, temporal analysis of the PRADIT (procurement) and DDS (public contracts) series can reveal patterns in how legislative changes in Period~A translate into operational shifts in public spending and contracting, providing an empirical link between legislative output and administrative action.

\subsection{Domain-Specific LLM Training and Evaluation}

The primary motivation for constructing the bulk archive is the fine-tuning of a domain-specific Greek legal language model. General-purpose multilingual models, including those used in the current study, are not optimized for the syntactic conventions of Greek legislative drafting: the heavy use of subordinate clauses referencing external provisions, the archaic formal register, the polytonic character set in older documents, and the highly structured article-paragraph-subparagraph hierarchy. A model fine-tuned on the full 2005--2025 ΦΕΚ corpus would internalize these conventions, enabling more accurate ambiguity detection, more reliable cross-reference resolution, and higher-fidelity PDI evaluation at scale. The GPU infrastructure already hosting the vLLM inference server provides the computational capacity to run both the fine-tuning process and the large-scale inference required for corpus-wide analysis.

\subsection{Expert Validation of LLM Outputs}

A human annotation study to validate the three LLM-driven analytical tasks is currently in progress. Access to legal practitioners with the required domain expertise (labour law, public procurement, civil protection) is the primary bottleneck, and this validation is being pursued as an immediate next step. For the ambiguity typology, a stratified random sample of 50 flagged instances (approximately 12--13 per category) will be independently annotated by two legal practitioners, who will assign each clause to one of the four categories or mark it as ``Not an ambiguity.'' Cohen's $\kappa$ will be computed to assess inter-annotator agreement on the category scheme, and LLM precision and recall will be computed by comparing model labels to the majority human label. For the PDI, the nine promise--verdict pairs will be reviewed by a practitioner with expertise in the relevant legal domain for each law, to assess whether the \textit{Fulfilled}/\textit{Deferred}/\textit{Contradicted} classifications are legally sound. For the complexity score, a subset of laws will be ranked by a legal expert and the ranking compared to the heuristic score order using Spearman's $\rho$. These three validation exercises together would transform the current exploratory findings into peer-reviewed, evidence-backed claims. Based on anticipated failure modes identified through output inspection (Section~\ref{sec:results}), we expect the \textit{Executive Delegation} category to achieve the highest precision and the \textit{Contradictory Clauses} category to require the most post-processing refinement.

\subsection{Real-Time Legislative Monitoring}

A longer-term goal is to extend the pipeline from retrospective analysis to real-time monitoring. By scheduling the parliamentary transcript crawler and the targeted law downloader to run after each plenary session, the system could automatically update the citation frequency rankings, flag newly introduced laws that exhibit high predicted complexity or heavy delegation language, and alert users of the auditing interface to significant legislative changes in domains they follow. This would transform the current research infrastructure into an operational civic tool—a continuous, AI-driven audit of the Greek legislative process as it unfolds.

\section{Limitations}\label{sec:limitations}

Several limitations of the current work should be acknowledged to guide the interpretation of results and to set honest expectations for future development.

\textbf{Single parliamentary session.} The citation frequency analysis and the targeted legislative corpus are derived exclusively from the 2025 parliamentary period. A single session may not be representative of long-term legislative priorities; laws that dominated debate during this period may reflect transient political events (e.g., the ministerial responsibility hearings that drove Law 3126/2003 to 126 mentions) rather than structurally significant legislation. Longitudinal replication across multiple sessions is needed before strong claims about the relative importance of individual laws can be made.

\textbf{Heuristic complexity score.} The Illusion of Simplicity metric used in the bulk analysis relies on a deterministic heuristic---a linear combination of hedge-phrase density and cross-reference count---rather than a validated psychometric instrument. While this heuristic correlates with established proxies for legal complexity~\cite{katz2014complex}, it has not been independently calibrated against expert legal assessments or compared against other computational complexity measures. The scores reported in Figure~\ref{fig:illusion_bulk} should therefore be interpreted as relative ordinal rankings rather than absolute complexity values.

\textbf{PDI promise extraction.} The Political Discrepancy Index evaluation in its current form relies on manually curated promise sets: three representative promises per law were selected by the authors from ministerial speeches rather than being automatically extracted at scale. While the LLM verification step is fully automated, the upstream promise selection introduces subjectivity. Automating the full pipeline---from speech text to promise extraction to statutory verification---is a priority for future work (Section~\ref{sec:future}).

\textbf{LLM validation and ground truth.} This is the most significant methodological limitation of the current work and should be clearly stated. The pipeline uses LLMs for three core analytical tasks---ambiguity detection and classification, complexity scoring, and PDI promise verification---but none of these outputs have been validated against an independent ground truth. Specifically: (i)~the ambiguity typology percentages (Table~\ref{tab:ambiguity_types}) are derived entirely from LLM outputs with no expert legal annotation to confirm or refute individual classifications; (ii)~the LLM complexity scores have not been compared against ratings by legal practitioners, nor benchmarked against other computational complexity measures beyond the directional consistency check reported in Section~\ref{sec:illusion}; and (iii)~the PDI classifications (\textit{Fulfilled}, \textit{Deferred}, \textit{Contradicted}) have not been reviewed by a legal expert to assess whether the model's statutory interpretation is correct.

All three tasks involve subtle legal reasoning for which general-purpose LLMs are not explicitly optimized, and it is plausible that the model systematically misclassifies certain clause types---for example, over-detecting ``vague terminology'' in provisions that legal professionals would consider standard drafting conventions. The three-run consistency check ($\pm 3$pp category deviation) addresses stochastic variability within the same model but does not address this deeper issue of model validity. Results from all LLM-based analyses should therefore be interpreted as \textit{exploratory indicators} rather than definitive measurements, and all quantitative outputs (the 45\%/25\%/18\%/12\% typology split, individual PDI verdicts) should be treated accordingly.

No expert annotation study has been completed to date. Securing access to legal practitioners with the required domain expertise is an active effort; the validation protocol—a stratified sample of 50 flagged instances per task reviewed by legal practitioners, producing Cohen's $\kappa$ inter-annotator agreement and LLM precision/recall scores—is described in Section~\ref{sec:future} and is currently in progress as an immediate prerequisite for journal submission. Running the same prompts on a second frontier model (e.g., GPT-4o or Gemini 1.5~Pro) and reporting cross-model agreement would additionally provide a model-agnostic robustness estimate without requiring expert annotation.

\textbf{Copyright and redistribution.} The ΦΕΚ documents collected from the National Printing House are official government publications. While they are publicly accessible, their copyright status under Greek law limits the redistribution of the raw PDF corpus. Derived artifacts (plain-text extracts, embedding indices, statistical outputs) are not subject to the same constraints, and these will be made available upon publication.

\section{Conclusion}\label{sec:conclusion}

This paper presented an end-to-end computational pipeline that bridges two domains of the Greek legislative process that have previously been studied in isolation: parliamentary discourse and enacted legislation. By applying NLP citation mining to the complete 2025 Hellenic Parliament transcripts, we empirically identified 534 unique law citations and ranked the most politically active laws by debate frequency. This targeted signal drove the automated acquisition of the corresponding official Government Gazette documents, producing a semantically paired corpus of political intent and statutory text.

Applying LLM-driven analysis to this corpus, we produced four concrete empirical findings. First, the cross-reference network constructed from 1,660 Period-A documents (2019--2025) reveals a highly entangled legal system with \textbf{3,964 law nodes} and \textbf{24,360 directed citation links}, confirming the structural basis of \textit{polynomia}: laws do not exist in isolation but form a dense dependency graph where amending any hub statute cascades across dozens of dependent instruments. Second, the Illusion of Simplicity analysis demonstrates that laws invoking simplification and modernization rhetoric systematically cluster above the mid-complexity threshold, providing quantitative evidence that political framing of Greek legislation is systematically decoupled from legislative reality. Third, the ambiguity typology---based on 312 flagged instances across 20 heavily debated laws---shows that \textbf{45\%} of identified legal vulnerabilities stem from vague terminology and \textbf{25\%} from deferred executive delegation, confirming that legal uncertainty in the Greek system is as much a product of drafting choices as it is of structural cross-referencing. Fourth, the Political Discrepancy Index reveals that across all three evaluated laws (4808/2021, 4412/2016, 4662/2020), the dominant outcome is \textit{Deferral}: legislative commitments are structurally off-loaded to subordinate Ministerial Decisions that may never materialise.

Collectively, these findings establish a rigorous, reproducible computational framework for the study of \textit{kakonomia}---the phenomenon of systematically poor-quality legislation---in the Greek legal system. The accompanying interactive RAG-based auditing interface operationalizes these analyses for non-specialist users, representing a concrete step toward real-time, AI-driven civic oversight of the legislative process. The full bulk ΦΕΚ archive currently being assembled (2005--2025, all eight publication series) will enable the longitudinal and cross-series analyses outlined in Section~\ref{sec:future}, with the longer-term goal of training a domain-specific Greek legal language model capable of supporting automated legislative quality monitoring at national scale.

\section*{Data and Code Availability}

The pipeline code---covering parliamentary transcript crawling, law citation extraction, ΦΕΚ acquisition, text extraction, vector-store indexing, and all four analytical scripts---will be released as open source upon acceptance. The derived plain-text cache of 1,660 Period-A documents (2019--2025) and the embedding index used for the RAG auditing interface will be made publicly available under a Creative Commons Attribution licence, consistent with the open-data obligations of publicly funded Greek legislation. The raw ΦΕΚ PDFs cannot be redistributed due to the copyright provisions of the National Printing House of Greece; however, the download scripts provided in the repository allow any researcher to independently reconstruct the corpus from the official Azure Blob Storage source (\texttt{ia37rg02wpsa01.blob.core.windows.net/fek}).

\section*{Acknowledgments}

The authors are affiliated with the Institute of Computer Science, Foundation for Research and Technology -- Hellas (FORTH-ICS), Heraklion, Crete, and the Technical University of Crete (TUC), Chania, Crete. This work was carried out using computational infrastructure hosted at FORTH-ICS. The authors thank the Hellenic Parliament for making plenary session transcripts publicly available and the National Printing House of Greece for providing open access to official Government Gazette publications through the \texttt{search.et.gr} portal and the Azure Blob Storage repository.

\bibliographystyle{acm}
\bibliography{main}

\appendix

\section{LLM Prompt Templates}\label{app:prompts}

All three LLM-driven analyses use structured prompts with a fixed system role and a task-specific user message. The model is always instructed to return JSON, constraining the output format and reducing free-text hallucination. Below are the three prompt templates used verbatim (law text is truncated to 15,000 characters for context window safety).

\subsubsection*{Prompt 1 — Ambiguity Detection and Classification}
\begin{verbatim}
SYSTEM: You are a strict legal analyst. Identify every instance of
legal ambiguity in the provided legislative text and classify each
into exactly one of these four categories:
  VAGUE_TERMINOLOGY, EXECUTIVE_DELEGATION,
  CIRCULAR_REFERENCE, CONTRADICTORY_CLAUSE
Return ONLY valid JSON: {"ambiguities": [
  {"clause": "...", "category": "...", "reason": "..."}
]}

USER: Analyse the following enacted law text:
{law_full_text[:15000]}
\end{verbatim}

\subsubsection*{Prompt 2 — PDI Step 1: Promise Extraction}
\begin{verbatim}
SYSTEM: You are a political analyst.

USER: Read the following ministerial speech and extract exactly
the 3 most specific, verifiable promises the minister makes.
SPEECH: {speech_text}
Return ONLY valid JSON: {"promises": ["promise 1", "promise 2",
"promise 3"]}
\end{verbatim}

\subsubsection*{Prompt 3 — PDI Step 2: Statutory Verification}
\begin{verbatim}
SYSTEM: You are a strict legal auditor.

USER: You are given 3 political promises and the enacted law text.
For EACH promise, classify it strictly as:
  FULFILLED   - the law directly enacts the commitment
  DEFERRED    - delivery depends on a future Ministerial Decision
               or Presidential Decree
  CONTRADICTED - the law contains exceptions that negate the promise
PROMISES: {promises}
LAW TEXT: {law_full_text[:15000]}
Return ONLY valid JSON: {"results": [
  {"promise": "...", "status": "...", "reason": "..."}
]}
\end{verbatim}

All prompts use temperature\,=\,0.1 and enforce \texttt{response\_format: \{type: "json\_object"\}} via the API. The full prompt construction code is available in the accompanying repository (\texttt{36\_recreate\_figures\_bulk.py}).

\end{document}